**Homodyne Photonic Tensor Processor exceeds 1,000-TOPS**

Lian Zhou[1,*], Kaiwen Xue[2,*], Yun-Jhu Lee[1*], Chun-Ho Lee[1*], Yuan Li[1*], Kiwon Kwon[2], Weipeng Zhang[1], Songlin Zhao[1], Jason Moraes[1], Niranjan Bhatia[2], Ryan Hamerly[1], Mengjie Yu[1,2], Zaijun Chen[1,2,#]

[1]*Opticore Inc., Berkeley, CA 94704*
[2]*Electrical Engineering and Computing Sciences, University of California, Berkeley, CA 94720*
**These authors contributed equally*
[#]*Author e-mail address: zaijun.chen@opticore.ai, zaijun@berkeley.edu*

## Abstract

High-performance computing underpins modern artificial intelligence (AI) [1], enabling foundation models, real-time inference and perception in autonomous systems, and data-intensive scientific simulations. Recent advances in quantization techniques [2,3] utilizing low-precision computation without degrading model accuracy, creates new opportunities for analog photonic computing [4–10] characterized by ultra-high clock rates and low energy consumption. Here we propose and demonstrate a coherent homodyne integrated circuit capable of general matrix multiplication (GEMM) with aggregate throughput that exceeds 1,000 TOPS (tera-operations per second), enabled by massive on-chip optical fanout and parallelism. By leveraging time multiplexing, the required modulator count is reduced from $O(N^2)$ to $O(N)$, allowing dense integration of record-scale 256 × 256 homodyne units (each <0.0064 $mm^2$) within a single reticle. We employ wafer-scale fabricated 64 thin-film lithium niobate (TFLN) transmitters (each over 40-GHz bandwidth with propagation loss of 0.2 dB/cm) to encode data and chip-to-chip coupled to Si/SiN computing circuits (64 channels). Our system achieves up to 7-bit computational accuracy across 8 × 8 parallel channels at record computing clockrate 120 Gbaud/s, and 6-bit statistical accuracy across 256 × 100 channels at 20-128 Gbaud/s, representing a total throughput of 1,000-6,000 TOPS. Massive parallelism amortizes the optoelectronic (OE) conversion to allow 330-TOPS/W efficiency using foundry-available packaging technology. The system throughput is benchmarked with Qwen2.5-0.5 billion parameter models that generate accurate tokens. High throughput and energy efficiency establish a near-term pathway toward light-based accelerators for large-scale training and low-latency inference from datacenters to edges, accelerating new models toward artificial general intelligence.

## Introduction

The rapid scaling of deep neural networks, particularly transformer-based [11] and diffusion models [12], has driven unprecedented advances in AI while placing increasing demands on computing hardware. In transformer-based large language models (LLMs) [13], an input query is first processed in a *prefill* stage, where the full context sequence is embedded and evaluated through a series of matrix multiplications, followed by an iterative *decode* stage that generates tokens sequentially (Fig. 1c). In practice, multiple user queries are processed concurrently, allowing both prefill and decode operations to be batched across requests. These workloads are dominated by general matrix multiplications (GEMMs) of the form $[L \times N] \times [N \times N]$ during prefill and $[B \times N] \times [N \times N]$ during decode, where $L$ is the context length, $N$ is the model dimension, and $B$ is the batch size across users. Recent trends show that context length has increased by orders of magnitude (Fig. 1b), directly scaling the size and frequency of these matrix operations. As a result, both the computational workload and data movement grow rapidly, with memory access becoming the dominant cost. In state-of-the-art electronic systems, these large-scale matrix operations are distributed across multiple accelerators and servers, requiring extensive data transfer between memory hierarchies and interconnects, which imposes significant energy and latency overhead, motivating alternative architectures that can execute large-scale matrix operations with higher parallelism and reduced data movement. Recent progress in model compression and quantization has shown that low-precision computation can preserve model accuracy across a wide range of workloads, enabling more efficient implementations of large-scale AI systems. These developments open new opportunities for alternative computing paradigms that operate at high throughput with reduced energy consumption.

Photonics offers new physics for computing by leveraging high bandwidth, low-loss data transmission, and intrinsic parallelism. Most current photonic architectures are with free-space optics [14–19] that allow high spatial parallelism but encounter challenges such as accuracy limits due to aberration-related crosstalk, speed limits due to capacitance for transmitters and receivers, as well as packaging solutions and compatibility with CMOS technology. Integrated photonic architectures based on spatial [4,6,9,10,20] and wavelength multiplexing [21,22] that rely on quadratic-scaling in modulator counts are limited in scalability due to chip area and device footprints. Time-multiplexed approaches [7,17,23–27], based on data mapping in amplitude-encoded optical pulses, reduce the number of modulators from $O(N^2)$ to $O(N)$, providing a potential to scale up optical systems. However, existing architectures suffer from complex $O(N^2)$ wavelength demultiplexing in cascade modulation [24,28], or phase instabilities using VCSELs [17], limiting their accuracy and system scalability.

Here we demonstrate (1) the record largest-to-date single photonic integrated compute circuit with 256x256 units, exhibiting system scalability; (2) the demonstrated highest-to-date computing clockrate at 128 Gbaud/s with 6-bit accuracy, 100-fold higher than digital electronics; and 7 bits at 40 Gbaud/s, representing 10-fold lower statistical errors than state-of-the-art performance [8]; (3) hybrid integration of thin-film lithium niobate (TFLN) and SiN/Si with 2 dB/coupling loss; (4) fully-packaged FPGA-controlled system parallel computing with 16 TFLN modulators coupling an 8x8 homodyne mesh with 8 bit accuracy at 2.4 Gbaud/s; (5) ultralow loss (<0.1 dB/channel), path-length symmetric design allows compact integration of 256×256 channels in a single photonic integrated circuits, representing a breakthrough in photonic circuit depth with record functional channel counts fabricated. A combination of high channel count (256×256), high clockrates (120 Gbaud/s), and high accuracy (6-8 bits) enables an unprecedented photonic computing throughput of 1000~10,000 TOPS in a single chip.

## System Architecture

Our computing architecture performs GMMM using space-time multiplexing $Y_{(N\times K)} = W_{(N\times M)}\ X_{(M\times K)}$ matrix operation with a N+K high-speed energy-efficient TFLN modulator array, and a N×K pathlength-matched homodyne detectors fabricated on the Si/SiN platform, as illustrated in Fig. 1c. The mesh is path length matched to have zero delay between input and weight vectors and split optical power equally through the crossbar.

The laser beam is spatially split to $N+K$ spatial beam paths to encode the input matrix $X_{(N\times M)}$ and the weight matrix $W_{(M\times K)}$ into optical pulses with $M$ time steps. Each path with a TFLN modulator draws data from memory (via a digital to analog convertor (DAC)) and converts it to optical pulses. In the Si/SiN photonic mesh, the $X$-beams and the $W$-beams are routed respectively, upward and downward along the diagonal directions. Such a pathlength matching scheme allows timing and phase matching to cancel phase instabilities due environmental temperature variations and fabrication errors. At the crossings, a portion of the $X$-beams and the $W$-beams is coupled out from the main waveguide and their interference results in a photon current $I_{n,k}$ proportional to the multiply–accumulate (MAC) product: $Y_{n,k} = \Sigma_M A^W_{m,n} \cdot A^X_{m,k} \cdot sin(\triangle\, \theta) \propto X_{m,n} W_{m,k}$, where $A^W_{m,n} = W_{m,n}$ and $A^X_{m,k} = X_{m,k}$ denote the amplitude of the $m^{th}$ time step in the n-th or k-th spatial mode (Fig. 1d). The integration is performed with an electronic capacitor in charge integrating amplifiers [17].

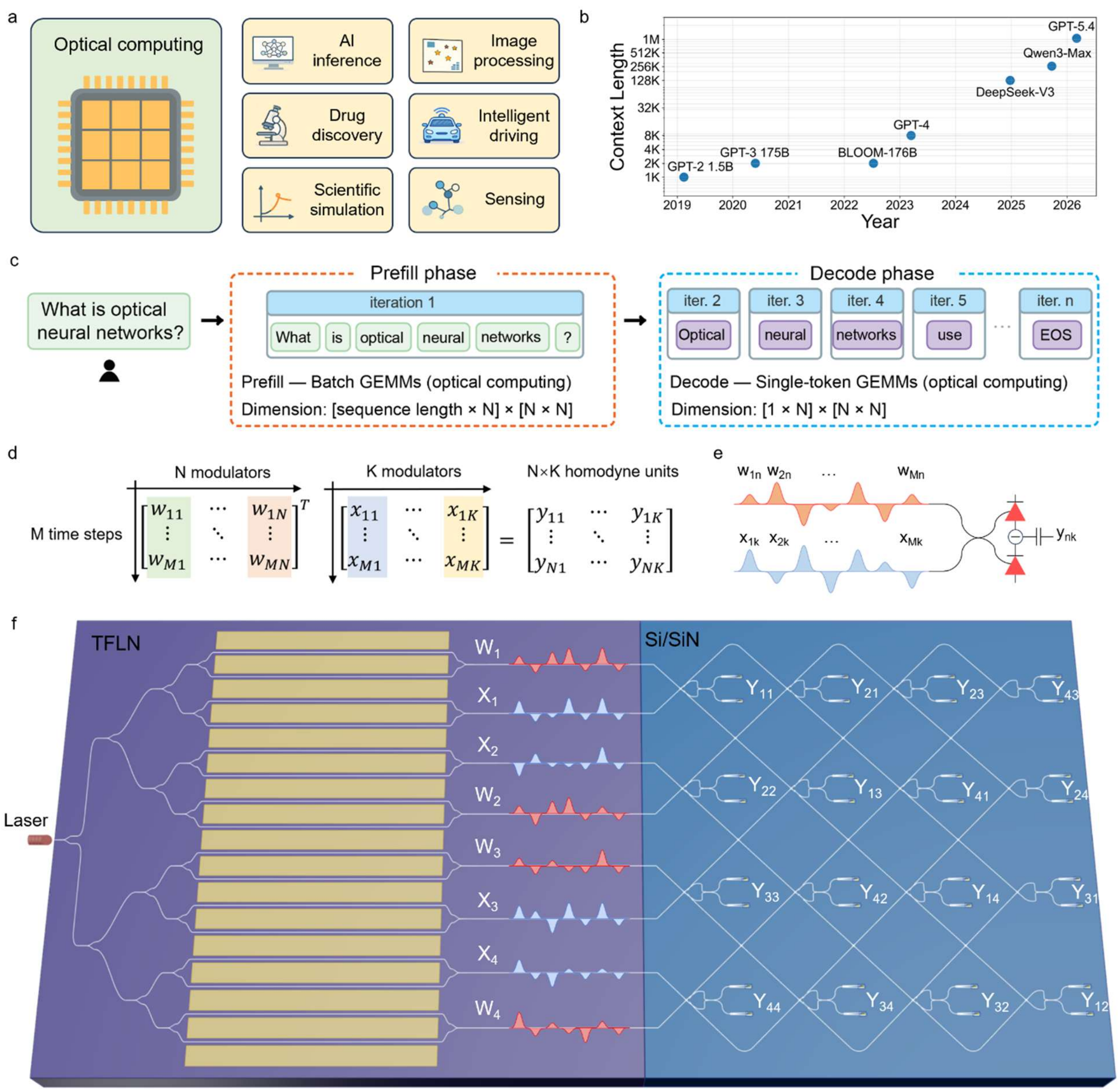

**Fig. 1|** Concept of TFLN–Si hybrid platform.
**a.** Application-level acceleration with optical computing.
**b.** The growth of computation in AI models.
**c.** LLM implementation accelerated by optical computing.
**d.** Matrix operation $Y_{(N\times K)} = W_{(N\times M)}\, X_{(M\times K)}$ with N+K modulators and N×K homodyne units.
**e.** Schematic of a single homodyne unit for vector multiplication.
**f.** Scalable photonics architecture for homodyne tensor processing.

## 2.2 Test results of Computing circuits

**8x8 computing crossbar circuits.**

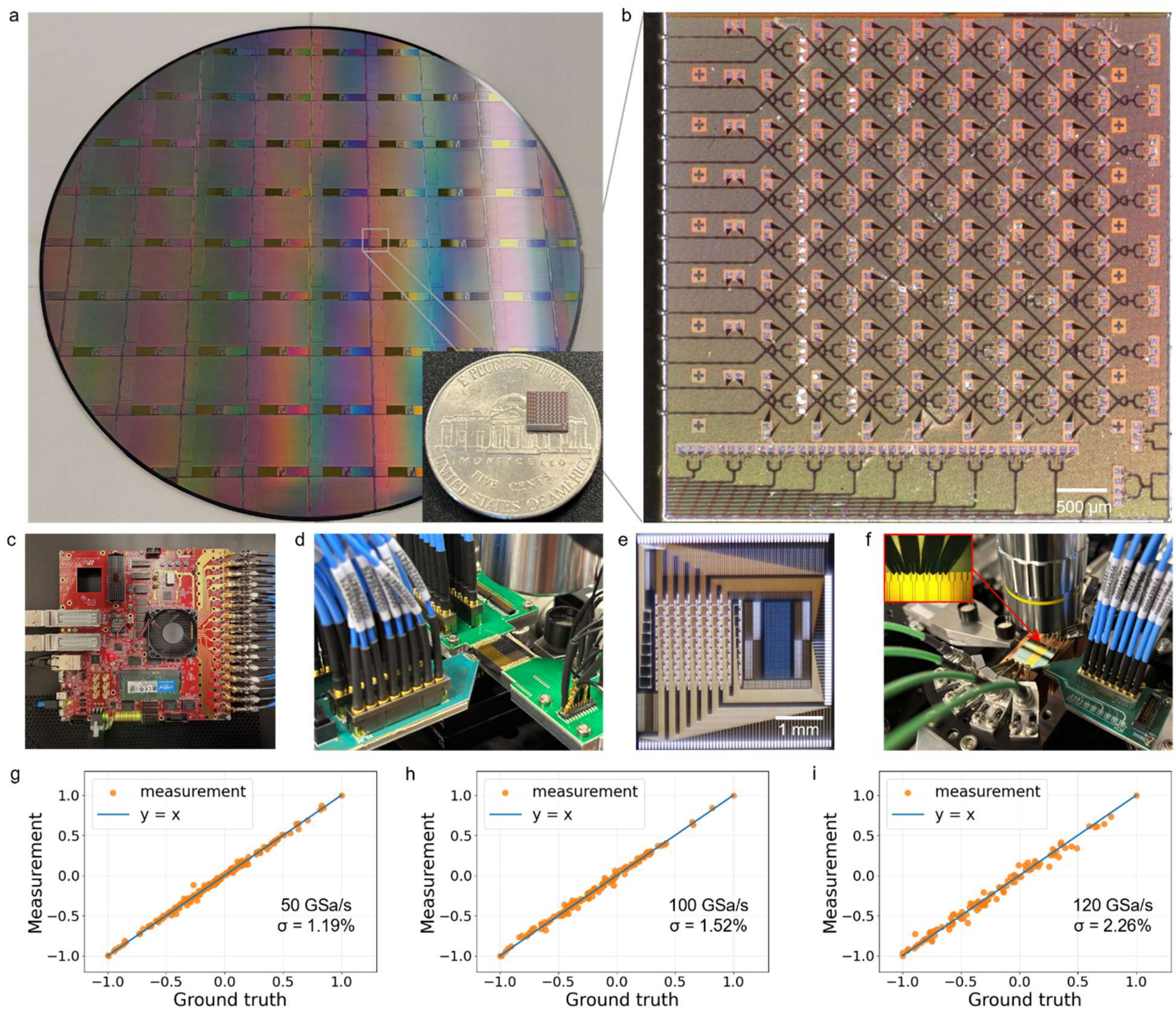

**Fig. 2|** Si/SiN homodyne crossbar packaging and benchmarking.
**a-b.** The 8×8 crossbar wafer and chip.
**c.** FPGA-based electronic subsystem.
**d.** Packaged TFLN-Si hybrid PIC platform.
**e.** Interposer for flip-chip bonding.
**f.** The setup with high-bandwidth multi-channel GSG probes for high-speed computing characterization.
**g-i.** Measured computing results of 7 bits at 50GSa/s, 6 bits at 100GSa/s, and 5.5 bits at 120GSa/s versus ground truth.

To demonstrate the capability of matrix operation, we integrate a 16-channel modulator array and a 8×8 homodyne mesh into the TFLN–Si hybrid photonic integrated circuit (PIC). On the TFLN chip, all electrical pads are routed to the chip edge to facilitate wire bonding through co-design of the bias heaters and RF electrodes. This layout also enables co-propagation of the optical and electrical signals, thereby supporting high-speed travelling-wave modulation. A 5 mm × 5 mm silicon photonic chip provides an 8×8 homodyne crossbar mapping 16 input channels into 64 parallel outputs (Fig. 2a-b). Wire bonding and microlens-assisted chip-to-chip edge coupling were employed to realize system-level interconnection and testing. To reveal the intrinsic high-speed computing capability, we construct a semi-packaged system using high-bandwidth GSG probes and high-speed AWG to replace the wire-bonded TFLN and FPGA (Fig. 2f). The modulator chip is driven at 120 GS/s, while the MAC signal is read out using the wire-bonded homodyne chip and a 3.5 MHz transimpedance amplifier (TIA). The measurement exhibits error deviations of 1.19% at 50 GSa/s, 1.52% at 100 GSa/s, and 2.26% at 120 GSa/s (Fig. 3h–j).

In parallel operations, the electronic subsystem is implemented by an FPGA (AMD RFSoC ZU49DR), which provides 14-bit ADC at 2.4 GSa/s, and real-time data processing (Fig. 2c). In the packaged TFLN-Si system (Fig. 2d), the FPGA feeds vectors with random numbers into the TFLN chip to modulate the laser amplitude over 16 channels, while a TIA array accumulates the differential photocurrent representing symbol-symbol multiplication to realize the MAC operation. The TIA array with tunable bandwidth converts high-speed photocurrent into low-speed voltage outputs for subsequent ADC readout and enables flexible accumulation for vectors of different lengths. Figure S1 in Supplementary Section S1 presents MAC results from 16 homodyne units, showing good agreement with the ground truth with a mean error deviation of 1.65%, corresponding to 7-bit accuracy. Following this systematic verification, the chip proceeds to a flip-chip packaging process with an interposer, which also enables reliable flip-chip bonding in future implementations (Fig. 2e)

**256x256 computing crossbar circuit**

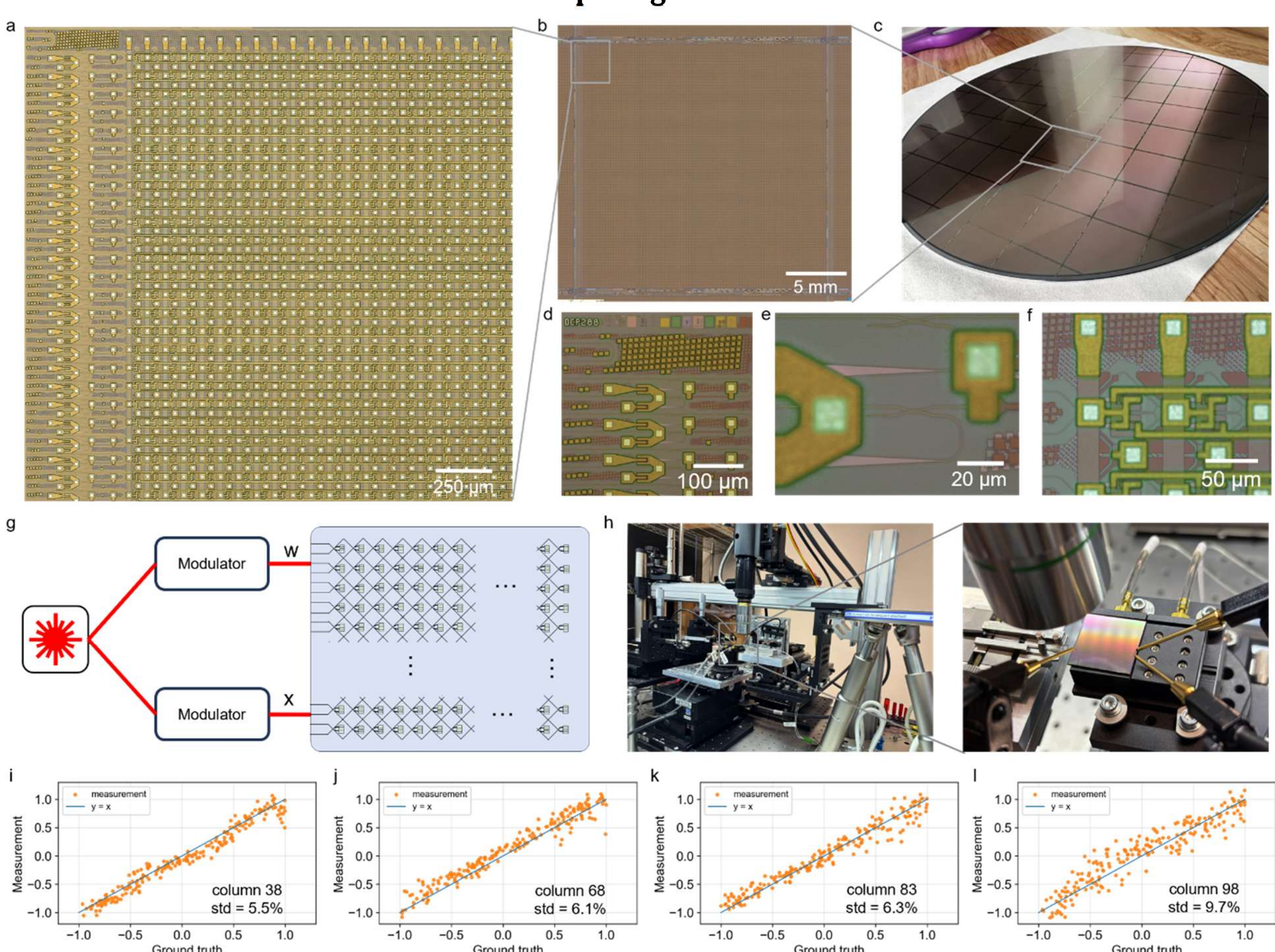

**Fig. 3|** Experimental characterization of a 256 × 256 homodyne chip.

**a-b.** Micrograph of the 256×256 chip and full-chip layout

**c.** Photograph of the wafer.

**d-f.** Zoomed-in views of the monitor PD on each input port, tap port, and homodyne units.

**g.** Schematic of the setup for calibrating the computational performance of the chip.
**h.** Automated chip characterization system using a motorized stage.
**i-k.** Measured computing results from homodyne units in column 38, 68, 83, and 98.

Figure 3 shows our large-scale 256 × 256 homodyne chip using a standard foundry process. As shown in Fig. 3a–f, the chip with size of 2x2 $cm^2$ features a dense homodyne array and representative zoom-in views of the device layout and structures. Each homodyne unit is individually calibrated by extracting the differential photocurrent from the balanced photodetector and maximizing it using an integrated phase shifter, thereby ensuring accurate and uniform readout across the array. With this architecture, the chip supports 65,536 parallel operations in each time step. To experimentally verify the homodyne performance of this large-scale array, we built a motorized measurement setup with two off-chip modulators and fiber arrays, as illustrated in Fig. 3g–h. Homodyne signals were obtained up to column 98 (Fig. 3i-l), confirming the operability of at least 256 × 98 matrix operation and 1000 TOPS at a clock rate of 20 GS/s. Although the error deviation remains relatively high due to phase fluctuations from the fiber-based modulators, these limitations can be mitigated by adopting a phase-stable on-chip modulator array. The upgraded platform is expected to enable high-speed and high-accuracy computing on the 256 × 256 homodyne chip, with a potential throughput of up to 15,728 TOPS at 120 GSa/s

**AI Inference on LLM and Resnet**

**Fig. 4|** AI model inference.
**a.** Dataflow of news title classification implemented with PIC.

**b.** Measured time trace and error histogram in stage B.
**c.** Fusion matrix of news title classification over 100 news samples.
**d-e.** Optical inference on CIFAR-10 and MNIST models.

To demonstrate the potential of the proposed platform for AI tasks, we implemented multiple inference models. Figure 4a presents an example of news-title classification performed on the TFLN–Si hybrid platform. Two stages requiring matrix operations, $G=Kr^TV$ ($Kr^T$ 128×64, V 64×64) and Y=QrG (Qr 64×128, G 128×64), were executed by the OPU. Figure 4b illustrates the measured time traces and error distribution in stage B (Y=QrG), exhibiting an error deviation of 0.53%. The optical inference achieves an accuracy of 78% over 100 test news titles, which is close to the 79% obtained in digital inference (Fig. 4c). For both CIFAR-10 and MNIST, the optical inference accuracy reaches 87% (vs digital 89.83% over 600 images) and 90.88% (vs digital 91.12% over 800 images) respectively, highlighting the accuracy and versatility of the platform across multiple AI tasks (Fig. 4d-e). Detailed model configurations and inference accuracy is illustrated in table 1.

Table 1. Comparison of digital and optical inference accuracy across different models

| Model configuration | Task | Samples | Digital accuracy | Optical accuracy |
|---|---|---|---|---|
| Multiple layers | MNIST | 800 | 91.12% | 90.88% |
| Convolution/Linear (ResNet) | CIFAR-10 | 600 | 89.83% | 87.00% |
| Transformer | News classification | 100 | 79% | 78% |

## System performance

The system throughput and energy efficiency is estimated for 256x256 channels at a clockrate of 20 GS/s, whereas 256x100 channels are characterized in existing circuits.
The total throughput T = 2x256x256x20 GS/s = 2,262.144 TOPS when packaging with drivers, nonlinearities, and memories at 1 pJ/access, which is achievable with existing foundry processes.
The total energy consumption for the system is calculated to be 8 W in Table 2. Therefore, the energy efficiency exceeds T/P=330 TOPS/W.

Table 2. Energy consumption for 256x256 channels, 20 GS/s, 8 bits

| Components | Energy budget | Amount | Total power |
|---|---|---|---|
| laser | 400 mW | 1 | 400 mW |
| Clock and Serdes (driver) | 2 mW | 512 | 1 W |
| Memory access | 0.5 pJ/data (4 bits) | 20 Gbaud/s<br>512 channels | 5.12 W |
| ADC @ 1 MHz | 0.01 mW | 256x256 | 0.66 W |
| Softmax and nonlinearity | 0.01 mW | 256x256 | 0.66 W |
| Total | | | 8 W |

**Data availability**

The data that support the findings of this study are available from the corresponding author upon reasonable request.

**Acknowledgement**

Z.C. and M.Y. acknowledge the DARPA NaPSAC program under project N66001-24-2-4002. Z.C. acknowledges discussion from Dr. Alex Sludds from MIT. Opticore acknowledges funding support from investors.

**Competing interests.**

Z.C. has filed a patent PCT/US24/50320 related to the technology in the work.

## Supplementary

## S1. 8×8 homodyne crossbar characterization

Figure S1 illustrates the MAC time traces through 16 homodyne units, which exhibits a mean std of 1.65% (7-bit accuracy). Each unit performs 100 integrations, each accumulating the products of random numbers.

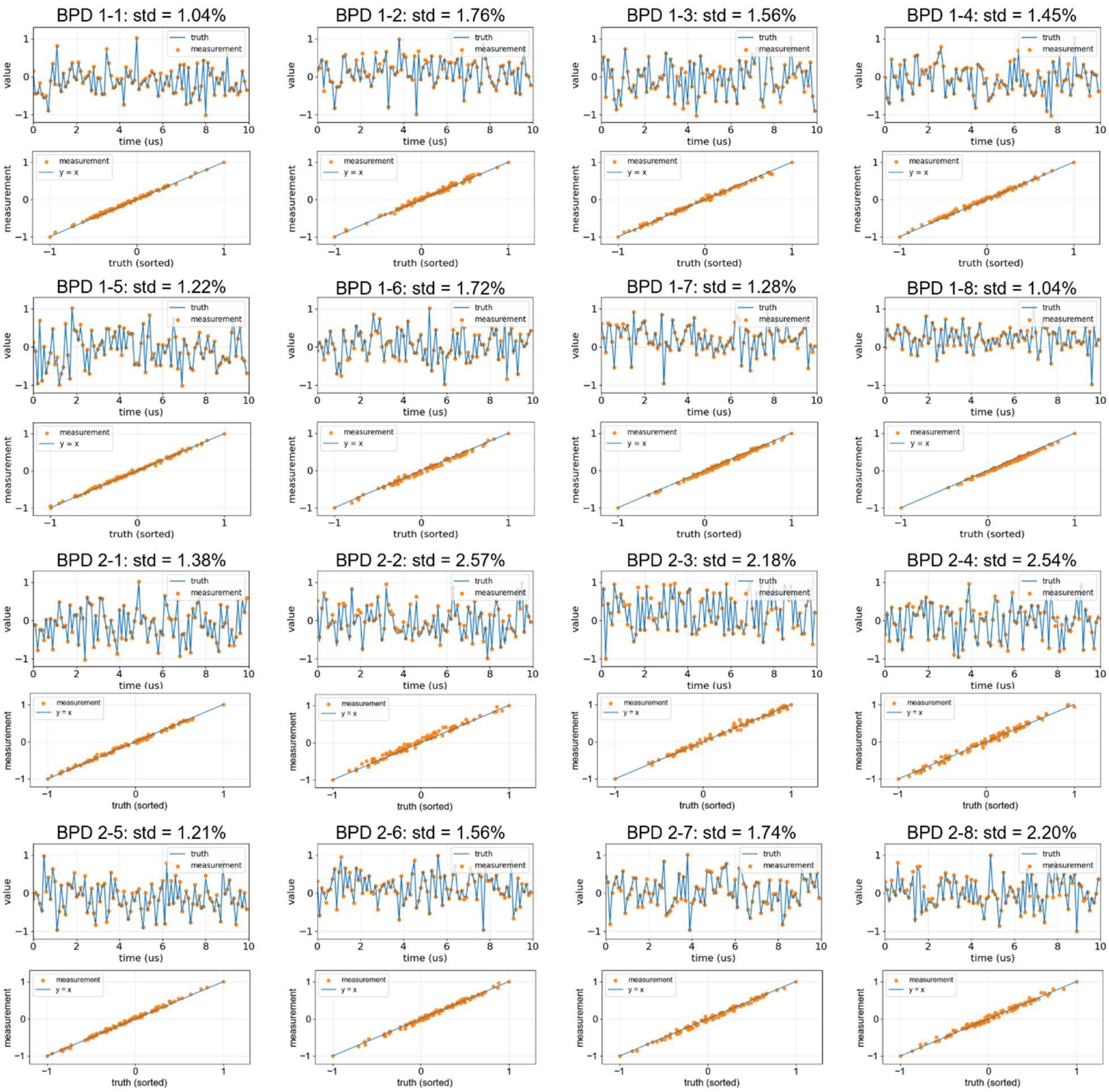

Figure S1. MAC measurement versus ground truth through 16 homodyne units